\documentclass[aps,pra,twocolumn,floatfix]{revtex4-1}

\usepackage{amssymb}
\usepackage{bm,dcolumn,amsmath}

\usepackage{float}
\usepackage{adjustbox}
\usepackage{graphicx}
\usepackage{amssymb,amsfonts,amsmath}
\usepackage{nicefrac}
\usepackage{color}
\usepackage{ulem} 

\usepackage{geometry}
\newcommand{\subf}[2]{%
  {\small\begin{tabular}[t]{@{}c@{}}
  #1\\#2
  \end{tabular}}%
}

\newcommand{\eref}[1]{Eq.~(\ref{#1})}

\begin{document}

\title{New ideas for tests of Lorentz invariance with atomic systems}
\date{\today}
\author{R. Shaniv$^{1}$}
\author{R. Ozeri$^{1}$}
 \author{M. S. Safronova$^{2,3}$}
 \author{S. G. Porsev$^{2,4}$}
 \author{V. A. Dzuba$^{5}$}
 \author{V. V. Flambaum$^{5}$}
 \author{H. H\"{a}ffner$^{6}$}
\affiliation{
$^1$Department of Physics of Complex Systems, Weizmann Institute of Science, Rehovot 7610001, Israel\\
$^2$Department of Physics and Astronomy, University of Delaware, Newark, Delaware 19716, USA\\
$^3$Joint Quantum Institute, National Institute of Standards and Technology and the University of Maryland, Gaithersburg, Maryland 20742, USA\\
$^4$Petersburg Nuclear Physics Institute, Gatchina, Leningrad District, 188300, Russia\\
$^5$School of Physics, University of New South Wales, Sydney 2052, Australia\\
$^6$Department of Physics, University of California, Berkeley, California 94720, USA}

\begin{abstract}
We describe a broadly applicable experimental proposal to search for the violation of local Lorentz invariance (LLI) with atomic systems. The new scheme uses dynamic decoupling and can be implemented in current atomic clocks experiments, both with single ions and arrays of neutral atoms. Moreover, the scheme can be performed on systems with no optical transitions, and therefore it is also applicable to highly charged ions which exhibit particularly high sensitivity to Lorentz invariance violation. We show the results of an experiment measuring the expected signal of this proposal using a two-ion crystal of $^{88}$Sr$^+$ ions. We also carry out a systematic study of the sensitivity of highly charged ions to LLI to identify the best candidates for the LLI tests.
\end{abstract}
\maketitle

Local Lorentz invariance (LLI) is a cornerstone of modern physics: the outcome of any local experiment is independent of the velocity  and the orientation of the (freely-falling) apparatus.
The field of Lorentz symmetry tests encompasses almost all fields of physics \cite{Mat05,LibMac09,KosRus11} and includes searches for Lorentz violation (LV) in the matter, photon, neutrino, and gravity sectors.
While the natural energy scale for strong LV induced by quantum gravity is the Planck scale ($M_{\mathrm{Pl}} \sim 10^{19}$ GeV/$c^2$), the  consequences of the Lorentz-violating physics may also lead to very small but potentially observable low-energy LV\cite{Kostelecky2004,Will2014}. Atomic physics LV tests were reviewed in \cite{SafBudDem17}. In this work, we develop new schemes and propose new systems for the LV tests in the electron-photon sector, performed with either trapped ions or neutral atoms using quantum-information enabled technologies, and provide proof-of-principle experimental demonstration.

LLI-violating effects are classified in the framework of the standard model extension (SME)~\cite{KosRus11,ColKos98}.
Violations of Lorentz invariance in
bound electronic states result in a small shift of the energy levels
described by a Hamiltonian~\cite{HohLeeBud13}
\begin{equation}
\delta H=-\left( C_{0}^{(0)}-\frac{2U}{3c^{2}}c_{00}\right) \frac{\mathbf{p}%
^{2}}{2}-\frac{1}{6}C_{0}^{(2)}T_{0}^{(2)},
\label{eq1}
\end{equation}%
where $\mathbf{p}$ is the momentum of a bound electron, $c$ is the speed of
light, and $U$ is the Newtonian gravitational potential.
The parameters $C_0^{(0)}$, $c_{00}$, and $C_{0}^{(2)}$
contain elements of the $c_{\mu \nu}$ tensor quantifying the LLI violation~\cite{HohLeeBud13,PruRamPor15}.
The relativistic form of the $T_0^{(2)}$ operator is $T_0^{(2)}= c \gamma_0 ({\boldsymbol \gamma} {\bf p} - 3 \gamma_z p_z)$,
where $\gamma_0$ and $\boldsymbol \gamma$ are the Dirac matrices.
The $c_{\mu \nu}$ tensor has nine components. The $c_{TJ}$  and $c_{TT}$  terms describe the
dependence of the kinetic energy on the boost of the laboratory frame and have a leading
order time-modulation period related to the sidereal year. The elements $c_{JK}$, where $J, K={X,Y,Z}$, describe the dependence of the kinetic energy on the direction of the
momentum and have a leading order time-modulation period related to the sidereal day (12~h and
24~h modulation).

The most sensitive LLI tests for electrons have been conducted with neutral Dy atoms
\cite{HohLeeBud13} and Ca$^+$ ions \cite{PruRamPor15}.
Recently, it was proposed to test LLI using a pair of two entangled trapped Yb$^+$  ions in the
$4f^{13}6s^2$~$^2F_{7/2}$ state of Yb$^+$ with the prospect to improve the current most stringent bounds by $10^5$ \cite{DzuFlaSaf16}.
However, the proposal of \cite{DzuFlaSaf16} requires using a decoherence-free subspace to cancel out magnetic field fluctuations. The need to prepare an entangled superposition of two ions, leads to three major difficulties: (1) applying it to the single trapped-ion clock experiments leads to a significant loss of sensitivity, (2) scaling it to a larger number of ions requires  creating a large number of entangled pairs, and (3) the scheme cannot be readily applied to highly charged ions which often lack strong optical transitions. The scheme proposed here mitigates all these problems without significant loss of sensitivity and provides a pathway to significantly extend the ultimate accuracy of LV tests in the electron-photon sector.
We also explore a possibility to use  highly charged ions or optical-lattice clocks to test the local Lorentz invariance violation and demonstrate enhancements of the LV violating effects in comparison with Yb$^+$.

\paragraph*{Experimental proposal.}
We  describe the proposed experimental scheme for the general case and  use the example of Yb$^+$ $^2F_{7/2}$ state for modeling.
The matrix element of the $T^{(2)}_{0}$ operator in Eq. (\ref{eq1}) is
\begin{eqnarray}
\label{eq10}
\langle J,m|T^{(2)}_{0}|J,m \rangle &=& \frac{-J\left(  J+1\right)  +3m^{2}}{\sqrt{\left(
2J+3\right)  \left(  J+1\right)  \left(  2J+1\right)  J\left(  2J-1\right)}} \,\nonumber \\ &\times&
\langle J||T^{(2)}||J \rangle,
\end{eqnarray}
where $J$ and $m$ denote the quantum numbers of the total electronic angular momentum and its projection on the quantization axis. Therefore, the tensor LV-violating signal is proportional to $m^2$. Thus, the experimental goal is to monitor the splitting  between different $m$ levels as the Earth rotates around its axis and around the Sun, and thus place a bound on $C_{0}^{(2)}$. Typically, the main source of decoherence in this type of experiments is the magnetic field noise leading to uncontrolled Zeeman shifts. In order to reduce the effect of magnetic field noise while maintaining the $m^{2}$ dependent effects, we propose a dynamical decoupling (DD) \cite{lidar2013quantum} technique that is applicable to spins of arbitrary size. \par

\paragraph*{General physical system description.}
We consider a spin $J$ system whose associated magnetic moment $\mu_{z}$ interacts with a magnetic field , $\textbf{B} = B_{z}\hat{\textbf{z}}$.  The Hamiltonian   $\mathcal{H}_{\mathrm{lin}} = \mu_{z}B_{z}J_{z}$ has equidistant energy eigenstates $\left|J,m\right>$. In addition to this linear Zeeman effect, we assume a small energy shift proportional to $m^2$, which can result from possible Lorentz violating terms but also from second order Zeeman shift or the electric quadrupole shift originating in ion traps from their inherent electric field gradient. This shift enters the Hamiltonian as $\mathcal{H}_{\mathrm{quad}} = \kappa J_{z}^2$. The total free evolution Hamiltonian is the sum of linear and the quadratic terms $\mathcal{H}_{\mathrm{free}} = \mathcal{H}_{\mathrm{quad}}+\mathcal{H}_{\mathrm{lin}} = \kappa J_{z}^{2} + \mu_{z}B_{z}J_{z}$.
We assume that we can drive our system with a radio-frequency (RF) oscillating magnetic field tuned close to the resonance transition frequency $\omega_{RF} = \frac{\mu_{z}B_{z}}{\hbar} + \delta\left(t\right)$, where $\delta\left(t\right)$ accounts for drifts in the ambient magnetic field at the spin's position. This drive translates to adding the time dependent coupling term $\mathcal{H}_{\mathrm{coup}}=\Omega\left(t\right)\cos\left(\omega_{RF}t+\phi\right)J_{x}$ to the Hamiltonian, where $\Omega$ is the multi-level Rabi frequency and $\phi$ is the RF phase. Moving to the interaction picture with respect to the oscillating magnetic field and applying the rotating wave approximation, we obtain the evolution Hamiltonian:
\begin{equation}
\mathcal{H}=\delta\left(t\right)J_{z}+\kappa J_{z}^{2}+\Omega\left(t\right)\left[J_{x}\cos\left(\phi\right)-J_{y}\sin\left(\phi\right)\right].
\label{Hamiltonian}
\end{equation}
In what follows, we assume that $\Omega\left(t\right)$ can take values of  $\Omega_{0}\gg\kappa,\delta\left(t\right)$ and $0$. According to Eq. (\ref{Hamiltonian}) that means that while applying a RF drive with duration $\sim \frac{\pi}{\Omega_{0}}$ the evolution due to $\mathcal{H}_{\mathrm{free}}$ can be neglected while the evolution due to $\Omega_{0}\left[J_{x}\cos\left(\phi\right)-J_{y}\sin\left(\phi\right)\right]$ is significant. \par
\paragraph*{Experimental scheme.}
In the following, we describe the dynamical decoupling method aimed at measuring $\kappa$ while mitigating the unwanted magnetic field noise $\delta\left(t\right)$ by a periodic modulation of $\Omega$ and $\phi$. This method is premised on a scheme published in Ref \cite{shaniv2016atomic} where it was used to measure the electric quadrupole shift, and is in a sense a generalization of the ubiquitous spin-echoed Ramsey sequence for a large spin $J$. For clarity, we describe a specific DD sequence although other types of DD sequences may be applied as well. The sequence begins with initializing our spin state in a specific $J_{z}$ eigenstate $\left|J,m=m'\right>$. A resonant RF pulse is then applied for a duration of $\tau=\frac{\pi}{2\Omega_{0}}$ ($\frac{\pi}{2}$ pulse). We define the phase of this pulse to be $\phi = 0$, and therefore the corresponding evolution operator is $\exp{\left(i\frac{\pi}{2}J_{x}\right)}$. This pulse maps the spin state to the corresponding $J_{y}$ eigenstate, and thus acts as the first $\frac{\pi}{2}$ pulse of a Ramsey sequence. Next, a modulation sequence is applied, in the form of \\ \centerline{[$t_{w}$]--[$\pi_{+y}$]--[$2t_{w}$]--[$\pi_{-y}$]--[$t_{w}$]}\\
where $\pi_{\pm y}$ are RF pulses with duration $\frac{\pi}{\Omega_{0}}$ ($\pi$ pulses) with $\phi=\pm \frac{\pi}{2}$ and $2t_{w}$ is the wait time between pulses, where the spin evolves freely. We choose the time $t_{w}$ such that over $4t_{w}$ time $\delta\left(t\right)$ changes slowly, and is effectively constant. Therefore, we can write the evolution of the spin system as,
\begin{eqnarray}
&&\mathcal{U}=\exp\left(i\left[\delta t_{w}J_{z}+\kappa t_{w}J_{z}^{2}\right]\right) \nonumber \\
&&\exp\left(-i\pi J_{y}\right)\exp\left(i\left[2\delta t_{w}J_{z}+2\kappa t_{w}J_{z}^{2}\right]\right)\exp\left(i\pi J_{y}\right)
\nonumber\\
&&\exp\left(i\left[\delta t_{w}J_{z}+\kappa t_{w}J_{z}^{2}\right]\right).
\label{DD_evolution}
\end{eqnarray}
As a result of the commutation relation $\left[J_{z}^{2},\exp\left(\pm i\pi J_{y}\right)\right]=0$, the signal term, $\kappa J_{z}^{2}$, generates a phase shift which is coherently accumulated during the sequence. However, $\left[J_{z},\exp\left(\pm i\pi J_{y}\right)\right]\ne0$, and therefore the phase due to the magnetic noise term $\delta\left(t\right)J_{z}$ is largely reduced by averaging. From a geometric point of view, the operation $\exp\left(\text{-}i\pi J_{y}\right)A\exp\left(i\pi J_{y}\right)$ acts as a $\pi$ rotation of the operator $A$ around the $\hat{y}$ axis. Such a rotation transforms $J_{z}$ to $\text{-}J_{z}$ and therefore the middle term in Eq. (\ref{DD_evolution}) equals to $\exp\left(i\left[\text{-}2\delta t_{w}J_{z}+2\kappa t_{w}J_{z}^{2}\right]\right)$.
Therefore the evolution operator $\mathcal{U}$ in the slow varying $\delta\left(t\right)$ approximation becomes $\mathcal{U}=\exp\left(i4\kappa t_{w}J_{z}^{2}\right)$ and the phase due to the linear Zeeman effect cancels.

Following $n$ repetitions of $\mathcal{U}$, a second $\frac{\pi}{2}$ pulse is applied, with an RF phase $\phi$ with respect to the first $\frac{\pi}{2}$ pulse. The evolution of the entire sequence, after a total time of $T=4nt_{w}$, can be written as,
\begin{eqnarray}
&&\mathcal{U}_{total}= \exp\left(\frac{\pi}{2}\left[J_{x}\cos\left(\phi\right)-J_{y}\sin\left(\phi\right)\right]\right) \nonumber \\
&& \exp\left(i\kappa T J_{z}^{2}\right) \exp\left(\frac{\pi}{2}J_{x}\right).
\label{total_evolution}
\end{eqnarray}
The phase $\phi$ of the last $\frac{\pi}{2}$ Ramsey pulse can be used to account for any systematic constant imbalance between wait times, that could arise from experimental imperfections.

\begin{figure}[h]
\begin{tabular}{|c|c|}
\hline
\subf{\includegraphics[width=0.24\textwidth]{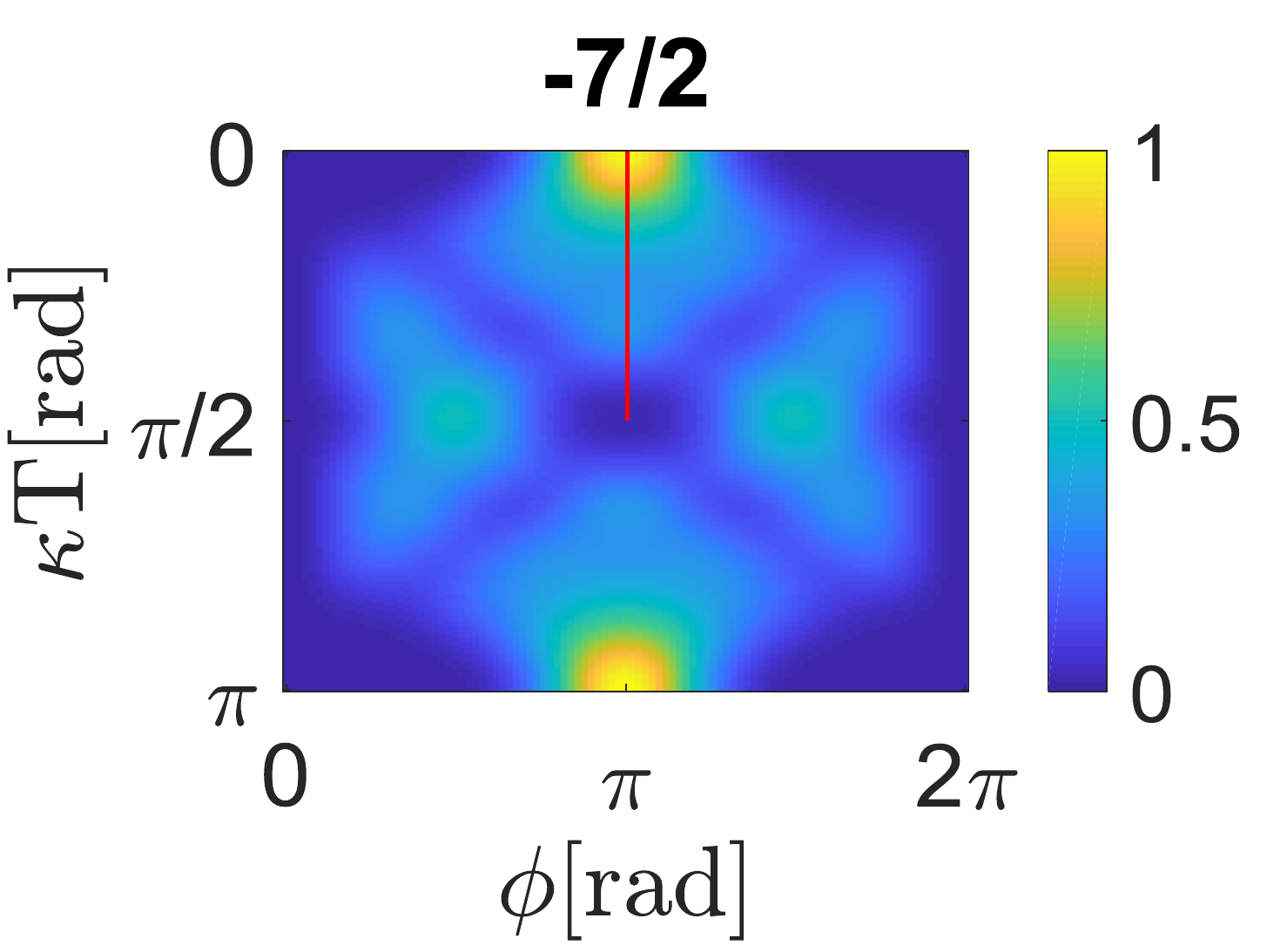}}{a}
&
\subf{\includegraphics[width=0.24\textwidth]{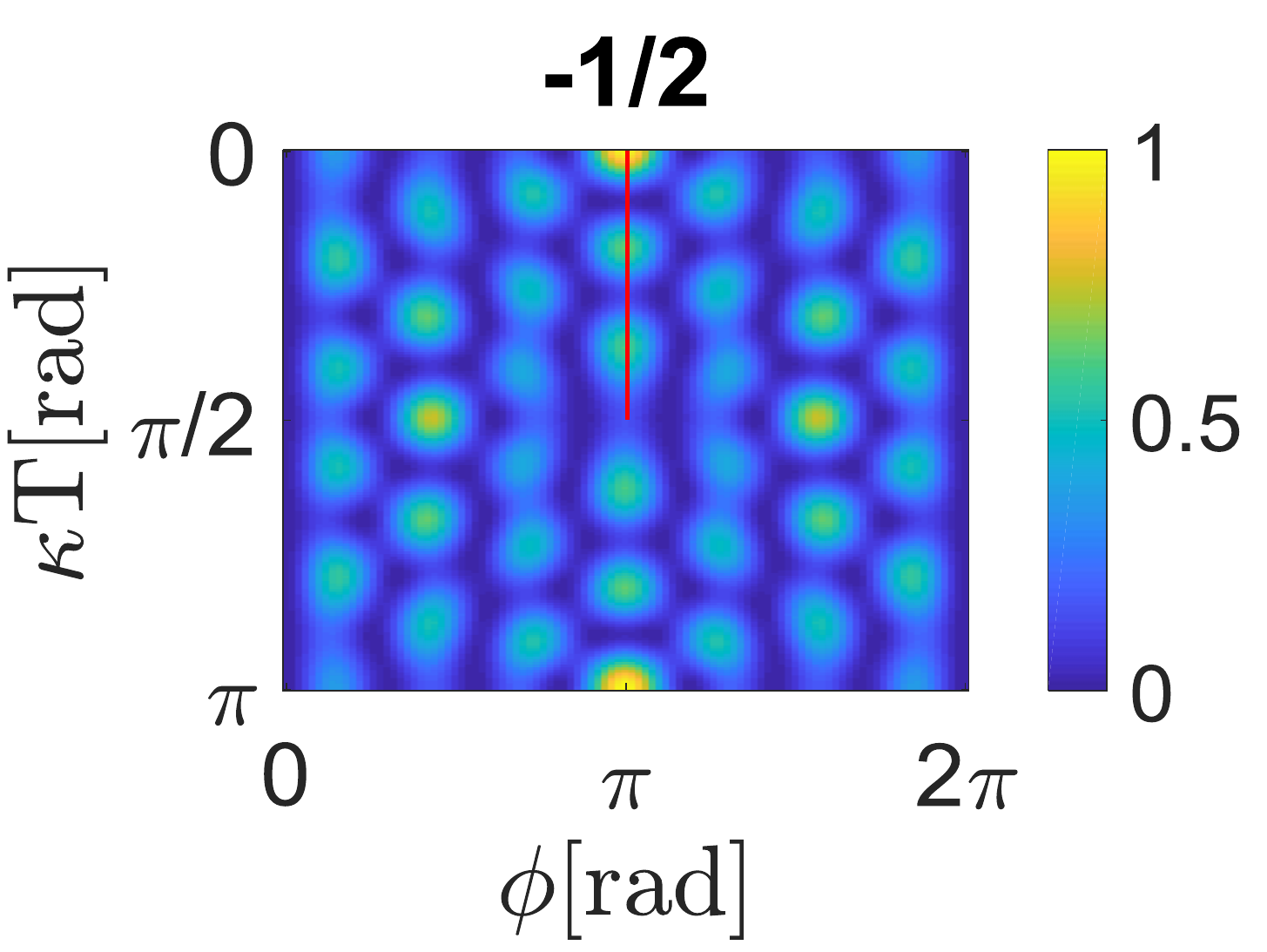}} {b}
\\
\hline
\subf{\includegraphics[width=0.24\textwidth]{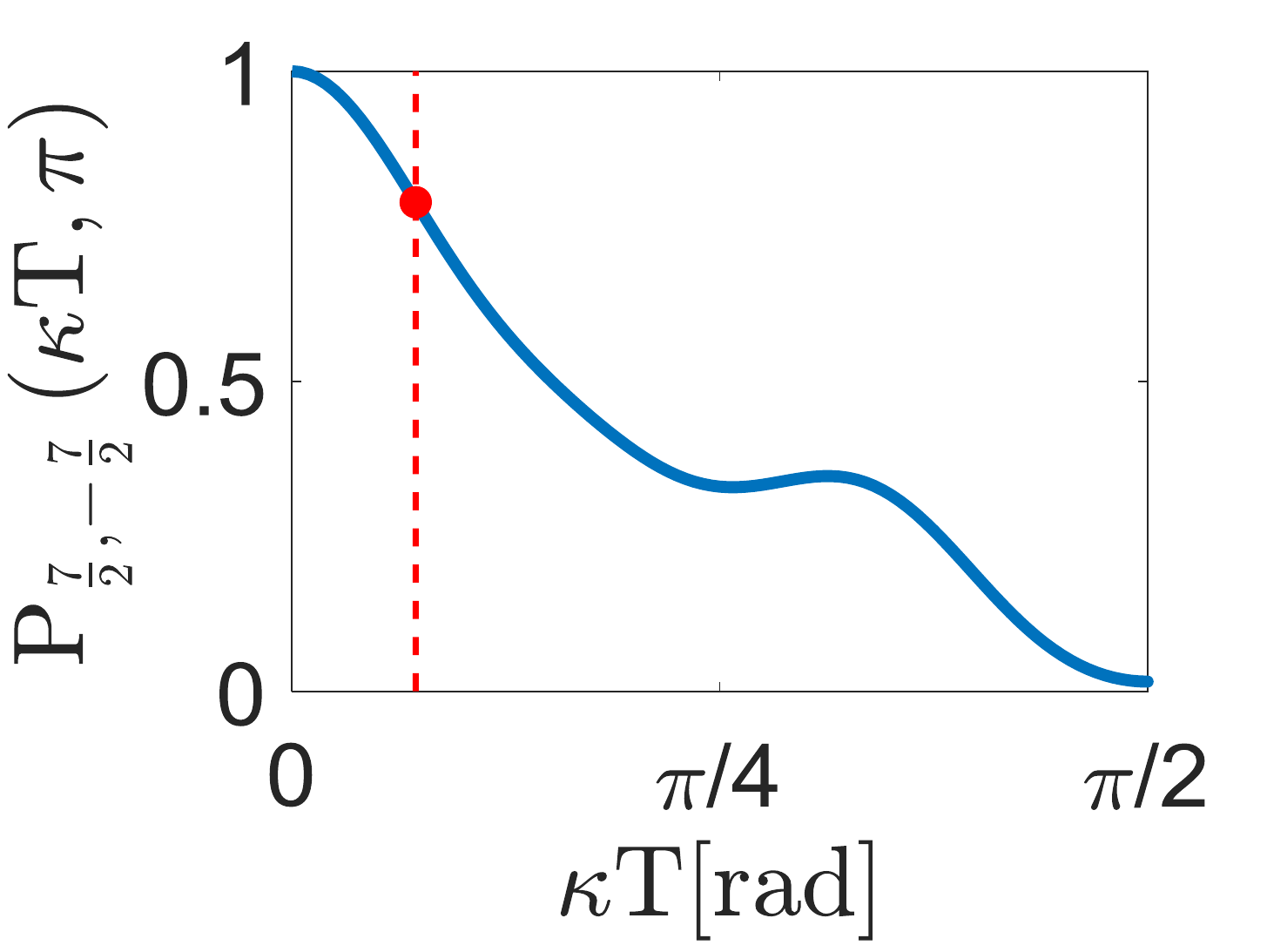}} {c}
&
\subf{\includegraphics[width=0.24\textwidth]{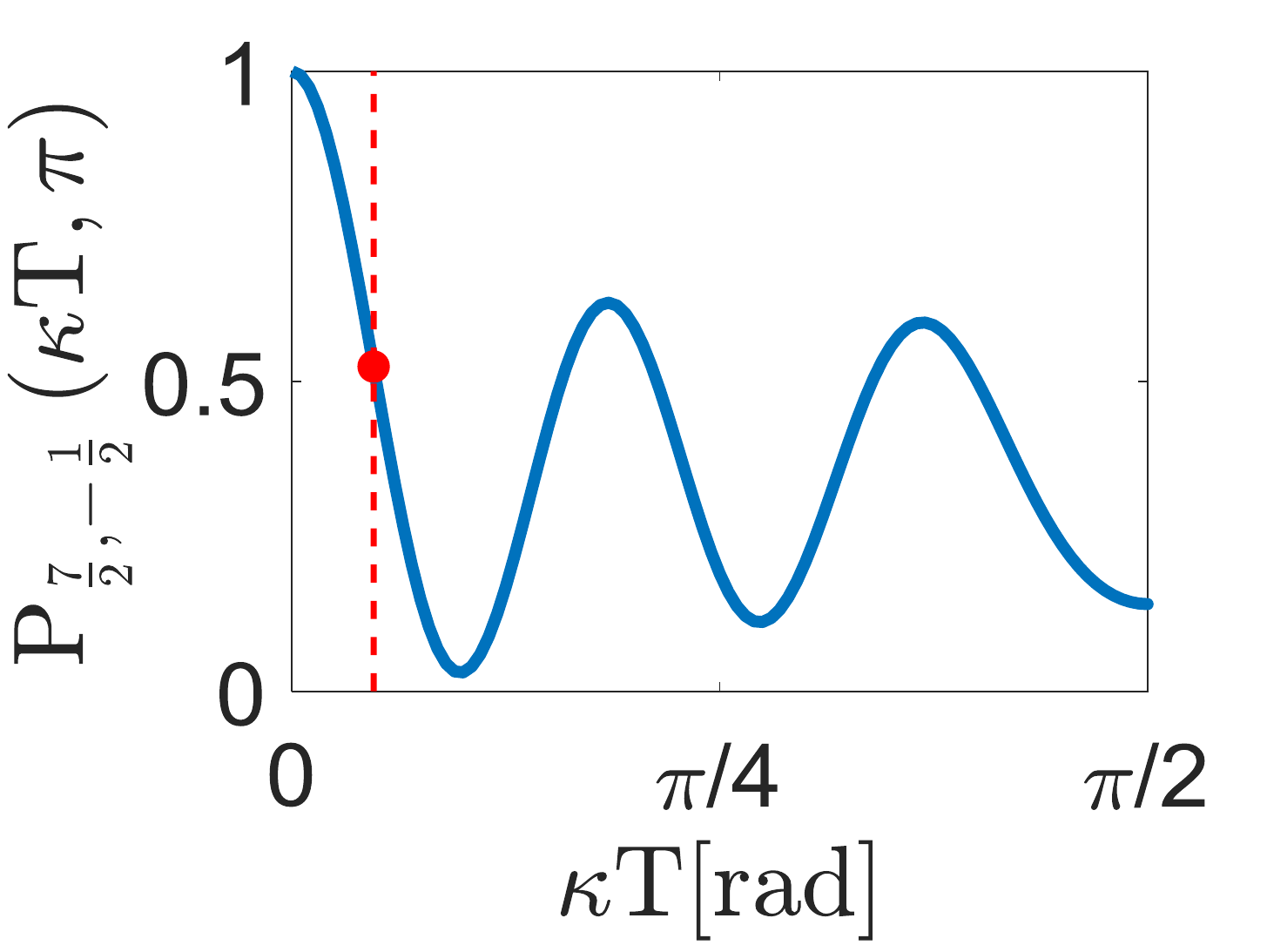}} {d}
\\
\hline
\end{tabular}
\caption{Theoretical calculation of $P_{\frac{7}{2},m}\left(\kappa T,\phi\right)$ for different $m$ values. $P_{\frac{7}{2},m}$ is periodic in $\kappa T$ with period of $\pi$ and it is symmetric with respect to $\pm m$. Therefore we only plot negative $m$ values and $\kappa T\in \left[0,\pi\right]$. \textbf{(a,b)} theoretical calculation of $P_{\frac{7}{2},\text{-}\frac{7}{2}}\left(\kappa T,\phi\right),P_{\frac{7}{2},\text{-}\frac{1}{2}}\left(\kappa T,\phi\right)$ as a function of $\phi$ and $\kappa T$ respectively.  Solid red line marks the $\phi=\pi$ line where the Ramsey fringe should be measured for maximal sensitivity. \textbf{(c,d)} Ramsey fringe in the $m=\text{-}\frac{7}{2},\text{-}\frac{1}{2}$ respectively, as a function of $\kappa T$. The curves correspond to the populations along the red solid lines in the top left and top right plots respectively.  Red dashed line marks the highest sensitivity $\kappa T$, and the red full circle marks the corresponding value of $P_{J,m}\left(\kappa T,\phi\right)$.}
\label{theory_for_7/2}
\end{figure}

Finally, the population in the initial state  $\left|J,m=m'\right>$, $P_{J,m'}\left(\kappa T,\phi\right) = \left|\left<J,m'\right|\mathcal{U}_{total}\left|J,m'\right>\right|^{2}$, is measured. Since $T$, the total experiment time, is known and $\phi$ can be calibrated,  $P_{J,m'}\left(\kappa T,\phi\right)$ can be directly used to estimate $\kappa$. $P_{J,m}\left(\kappa T,\phi\right)$ is therefore an equivalent of the Ramsey fringe in this large-$J$ Ramsey-sequence generalization. The theoretical calculation of $P_{J,m}\left(\kappa T,\phi\right)$ for $J=\frac{7}{2}$ and $m=\text{-}\frac{7}{2},\text{-}\frac{1}{2}$ are shown in Fig. \ref{theory_for_7/2}. By repeating this measurement sequentially in time and recording $P_{J,m}\left(\kappa T,\phi\right)$,  $\kappa$ can be extracted. Fig.~\ref{theory_for_7/2}c,d show the expected signal
as a function of $\kappa T$ for $\phi = 0$. The proposed experiment consists of monitoring the results of sequential measurements in time of $P_{J,m}\left(\kappa T,\phi\right)$, and look for time-dependent variation at the theoretical sidereal day and sidereal year periods. An optimal point to search for variations in $\kappa$ would be around the point at which $P_{J,m}\left(\kappa T,\phi = 0 \right)$ has the steepest slope with respect to $\kappa T$, indicated by the red dashed lines in Fig~\ref{theory_for_7/2}a,b. See supplementary material for further discussion.
Experimentally it will be likely easiest to choose the total Ramsey time $T$ to maximize the slope, but also the trap frequency and magnetic field can be used to tune  $\kappa$ via the electric quadrupole or second-order Zeeman shifts.
The best state to initialize and detect sensitivity-wise is $m=\text{-}\frac{1}{2}$, since it has the steepest slope. However, if preparation and detection of $m=\text{-}\frac{1}{2}$ is experimentally difficult, as in the case of logic initialization and detection of a highly-charged ion with no optical transition, then the $m=\text{-}\frac{7}{2}$ can be used.
\begin{figure*}
\begin{tabular}{|c|c|c|}
\hline
\includegraphics[width=0.28\textwidth]{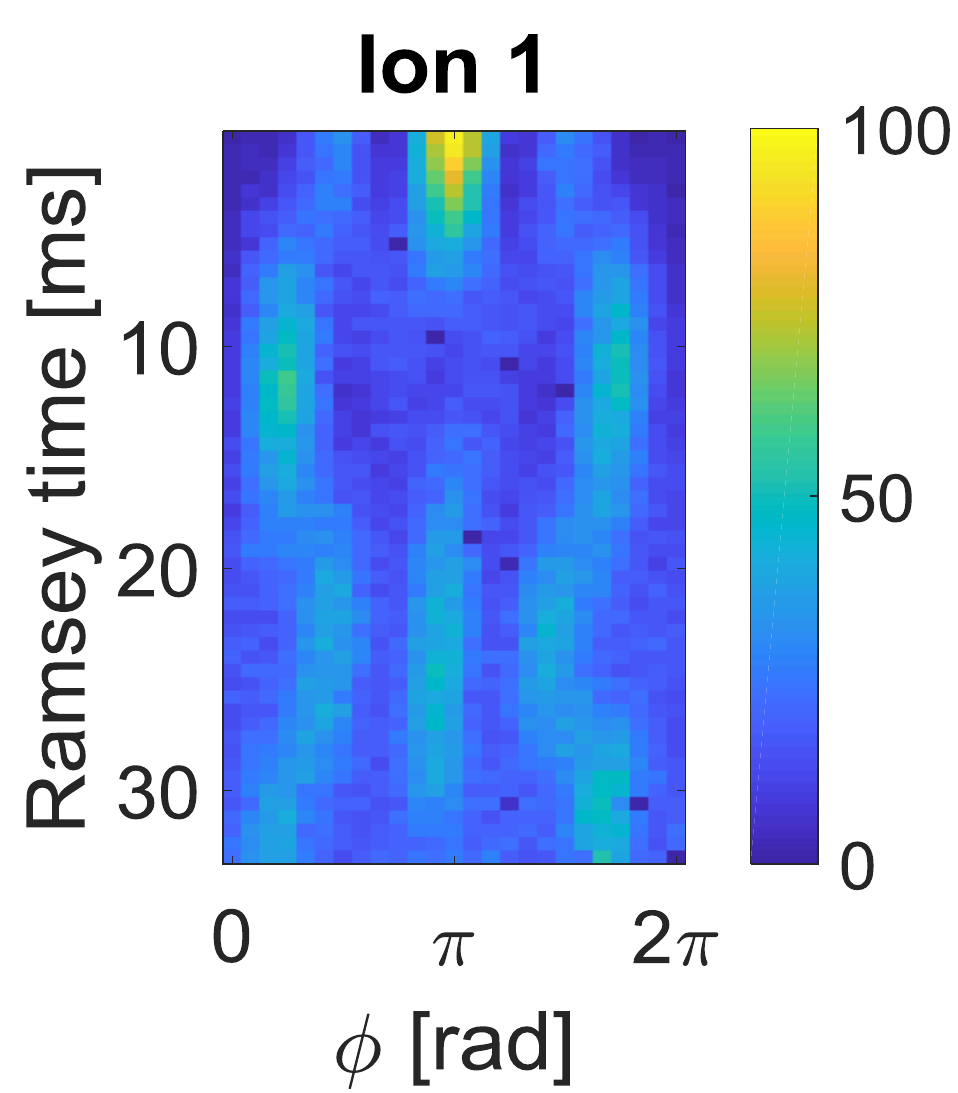}{a}
&
\includegraphics[width=0.28\textwidth]{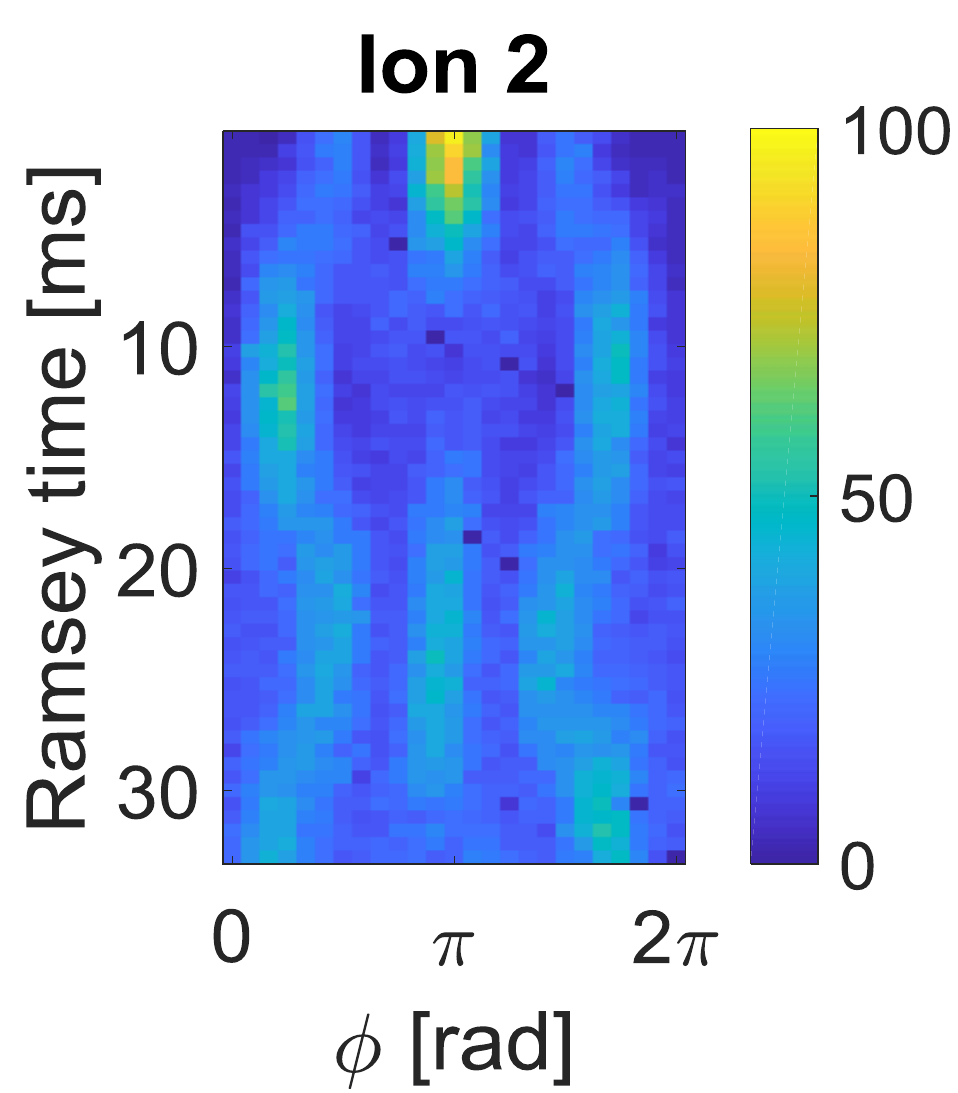}{b}
&
\includegraphics[width=0.28\textwidth]{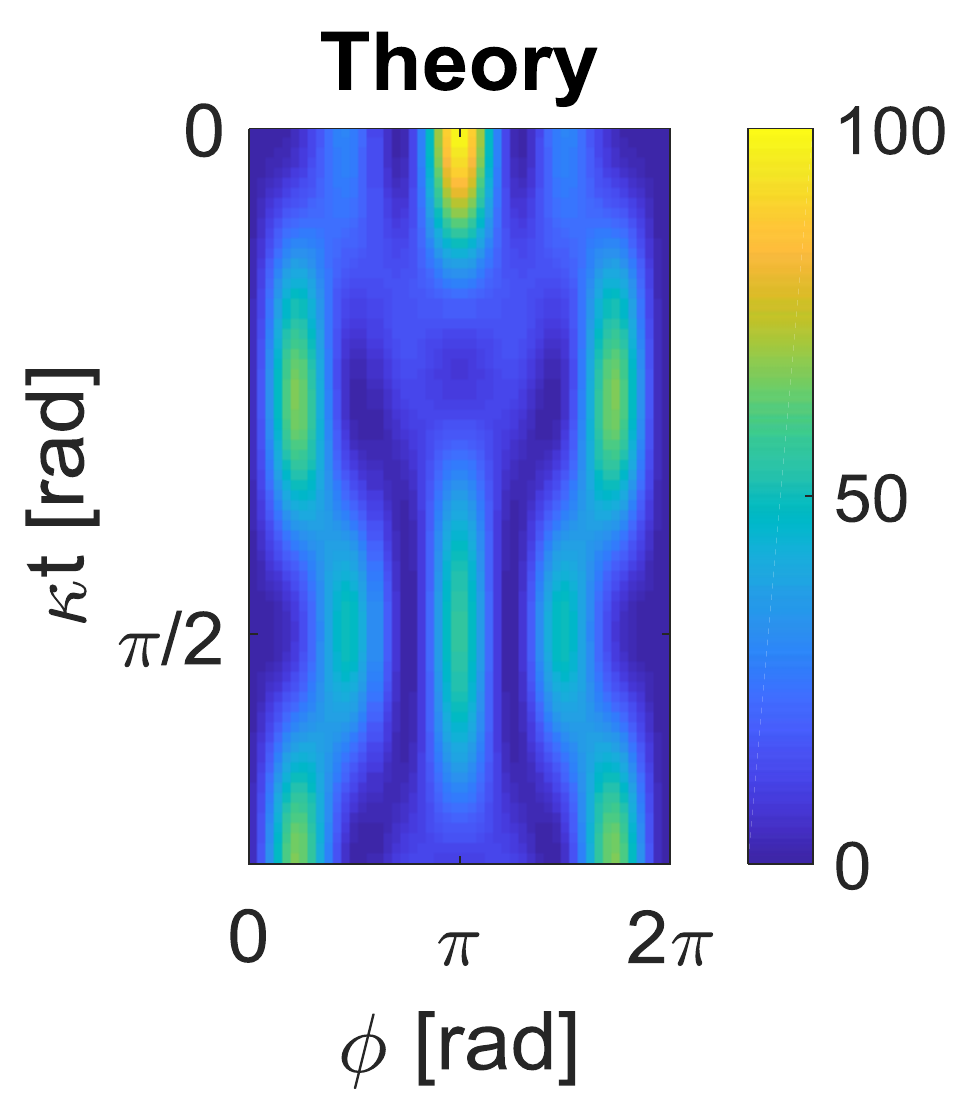}{c}
\\
\hline
\end{tabular}
\caption{Experimental verification of the DD method on the $4D_{\frac{5}{2}}$ level of two trapped $^{88}\mathrm{Sr}^{+}$ ions.  \textbf{(a,b)} Measurement of $P_{\frac{5}{2},-\frac{3}{2}}\left(\kappa t,\phi\right)$ in percent after the above DD sequence for different total experiment Ramsey time and $\phi$, for ion 1 and ion 2 respectively. In the experiment $t_{w}=150$ $ \mathrm{ \mu sec}$ and the DD pulse number goes from 2 to 110 (see supplemental material). \textbf{(c)} Theoretical calculation of $P_{\frac{5}{2},-\frac{3}{2}}\left(\kappa t,\phi\right)$.}
\label{experimental_data_for_5/2}
\end{figure*}
\par

Notice that this method enables the measurement of effects quadratic in $m$ while mitigating the effect of linear Zeeman shift noise by using only local spin operations. It is therefore straightforward to generalize this method for an ensemble of $N$ spins, e.g. a large ion chain or neutral atoms in an optical lattice. The uncertainty in evaluating $\kappa$ thus reduces by a factor of $\sqrt{N}$.\\
In addition, our procedure requires only initializing and detecting one specific state; $\left|J,m\right>$. This is useful in systems where logic spectroscopy \cite{schmidt2005spectroscopy} must be used, e.g. for highly charged trapped ions.  Moreover, even if weak optical transitions are required to initialize and read-out the final state, the coherent operations are carried out with RF only, thus avoiding effects from systematic AC-Stark shifts.\\

Another advantage of the RF-manipulation scheme is that the wavelength of the RF-radiation is much longer than the motional amplitudes of the ions allowing for high-fidelity coherent manipulation even at high temperatures. While one may still require optical fields to initialize and read out the states STIRAP or sequentially repeated pulses can be used yielding high state transfer fidelities even if the quality of a $\pi$-pulse would be low \cite{hume2007high}.
Finally, we note that one can also use strong RF field gradients to drive sideband transitions. As a consequence, one can apply quantum logic spectroscopy and detect the state of probe ions even if there are no optical transitions available opening up the possibility to use any HCI whose ground state has an angular moment of larger than 2$\hbar$.


\paragraph*{Measurement sensitivity.} One important aspect is how sensitive the presented method is as compared to the method presented in Ref.~\cite{DzuFlaSaf16}. The contribution  of Lorenz violation effects to $\kappa$ is given by Eq.~\ref{eq10}
\begin{equation}
\kappa_{\mathrm{LV}}/2\pi=5.1\times10^{15} \rm{Hz} \cdot C_{0}^{(2)}.
\end{equation}
In the supplement, we evaluate the measurement precision $\Delta \kappa$ with which $\kappa$ can be measured for $J=\frac{7}{2}$. We find that it is optimal to use $m=\frac{1}{2}$ as an initial state and estimate for this case $\Delta \kappa = 0.1\:\frac{\rm rad}{\sqrt{N\tau T}}$ where $T,\tau$ and $N$ are the interrogation time, total integration time and the number of spin probes, respectively. For comparison, $\Delta \kappa$ calculated for the method presented in Ref.~\cite{DzuFlaSaf16} is $\Delta \kappa = 0.083\:\frac{\rm rad}{N\sqrt{\tau T}}$. While for small ion or atom numbers $N$ both methods yield similar precisions, the method presented here can be readily extended to larger $N$, while the method in Ref.~\cite{DzuFlaSaf16} is more difficult to scale due to the complexity in exploiting quantum correlations.


\paragraph*{Proof-of-principle experimental demonstration.}

In order to verify the scheme, we measured $\kappa$ for the $4D_{\frac{5}{2}}$ level in two $^{88}$Sr$^{+}$ ion chain trapped in a linear Paul trap. The dominant contribution to $\kappa$ comes from the quadrupole shift, which can be used as a resource to tune our system to the most sensitive measurement point. We initialized our ions in the $m=-\frac{3}{2}$, and implemented the above DD sequence for times between $600$ $\mathrm{\mu sec}$ and $33$ $\mathrm{msec}$, with up to 110 pulses using $t_{w}=150$ $\mathrm{\mu sec}$. The results along with the corresponding theoretical expectations are presented in Fig.~\ref{experimental_data_for_5/2}.

\begin{table}[t]
\caption{The reduced matrix elements $|\langle J ||T^{(2)}|| J \rangle|$ (in a.u.) and LLI-induced energy shift (in Hz)  between the
highest and lowest values of $|m|$. The Ca$^+$, Yb$^+$, and Yb values are for the excited states, all other values are for the ground states.
$N$ is the number of the electrons in an ion. }
\label{ions}%
\begin{ruledtabular}
\begin{tabular}{lclccc}
Ion     &  $N$ &     Level     &  $J$&$|\langle J ||T^{(2)}|| J \rangle|$ & $|\Delta E/(h C_0^{(2)})|$\\
\hline      \\[-0.4pc]
  Ca$^{+}$   &19& $3d$          &5/2&    9.3   & $4.5 \times 10^{15}$~\cite{PruRamPor15} \\ Yb$^{+}$   &69& $4f^{13}6s^2$ &7/2&    135  & $ 6.1 \times 10^{16}$~\cite{DzuFlaSaf16} \\[0.3pc]
  Tm         & 69 & $4f^{13}6s^2$  &7/2& 141  &  $6.4 \times 10^{16}$ \\
  Yb         & 70 & $4f^{13} 5d 6s^2$ &2& 74   &   $3.9 \times 10^{16}$ \\[0.3pc]
  Th$^{3+}$  &87& $5f$          &5/2&    47  & $2.2 \times 10^{16}$  \\
  Sm$^{15+}$ &47& $4f$          &5/2&   128   & $5.7 \times 10^{16}$  \\
  Sm$^{14+}$ &48& $4f^2$        & 4 &    124  & $5.5 \times 10^{16}$ \\
  Sm$^{13+}$ &49& $5s^2 4f$     &5/2&    120  & $5.8 \times 10^{16}$ \\
  Eu$^{14+}$ &49& $4f^2 5s$     &7/2&    120  & $5.4 \times 10^{16}$ \\
  Nd$^{10+}$ &50& $4f^2$        & 4 &     96  & $4.3 \times 10^{16}$ \\
  Cf$^{15+}$ &83& $5f 6p^2$     &5/2&    112  & $5.4 \times 10^{16}$ \\
  Cf$^{17+}$ &81& $5f$          &5/2&    144  & $6.9 \times 10^{16}$ \\   [0.3pc]
 Os$^{18+}$ &58& $4f^{12}$& 6  &   367 & $1.4 \times 10^{17}$ \\
 Pt$^{20+}$ &58& $4f^{12}$& 6  & 412  & $1.6 \times 10^{17}$ \\
 Hg$^{22+}$ &58& $4f^{12}$& 6  & 459  & $1.8 \times 10^{17}$ \\
 Pb$^{24+}$ &58& $4f^{12}$& 6  & 507  & $2.0 \times 10^{17}$ \\
 Bi$^{25+}$ &58& $4f^{12}$& 6  & 532  & $2.1 \times 10^{17}$ \\
  U$^{34+}$ &58& $4f^{12}$&  6 & 769  & $3.0 \times 10^{17}$ \\
\end{tabular}
\end{ruledtabular}
\end{table}
\paragraph{Neutral atoms in optical lattices.}
Our DD scheme can also be applied to neutral atoms which allow for a large number $N$ of probes and have already been successfully employed for LV tests in electromagnetic sector \cite{HohLeeBud13}. To overcome systematic effects it may be advantageous to trap them in optical lattices where potentially $10^5$ or more atoms may be held in the future \cite{CamHutMar17}.
In the current lattice clocks, such as Sr, Yb, or Mg, $J=0$ states are used exhibiting no sensitivity to tensor LV in the electromagnetic sector.
Nevertheless, other precision LV tests could be possible with neutral atom clocks, such as for example measuring LV effects due to the first term in Eq.~(1) and measuring $c_{\mu \nu}$ in the nucleon sector using isotopes with nuclear spin $I>1/2$ see \cite{Wolf2006,Smiciklas2011,PihGueLas17,FlaRom17}.  For the LV tests in the electron sector with neutral atoms, the ground state of Tm, having the
 the same electronic $4f^{13}6s^2$~$^2F_{7/2}$ configuration as Yb$^+$, appears to be rather well suited as it has the same high sensitivity as Yb$^+$. Moreover, Tm is already being pursued for the lattice clock development, and trapping of the ensemble of Tm atoms
in a 1D optical lattice has been demonstrated \cite{SukFedTol16}. We note that a Tm clock is not needed for an LV test, just the ability to perform the scheme described here for the Tm ground state. Using Yb, the metastable  $4f^{13} 5d 6s^2$~$J=2$-state
could be used, too. For neutral atoms held in optical lattices, an additional systematic effect may arise from the trapping beams due to ac Stark shifts of the Zeeman components. 

\paragraph*{Highly charged ions.}

A number of highly charged ions (HCI) were recently shown to be  candidates for the  development of atomic clocks
and the search for variation of the fine-structure constant $\alpha$ \cite{BerDzuFla10,SafDzuFla14}.
Experimentally, sympathetic cooling of HCI was demonstrated in~\cite{SchVerSch15} for Ar$^{13+}$ and the spectra of
Ir$^{17+}$ ion, suitable for the above applications, were explored in Ref.~\cite{WinCreBek15}.
We have carried out the calculation of the matrix elements of the $T_0^{(2)}$ operator in the wide range of HCIs and find enhancement in the LV effects for the states containing 1-2 valence electrons or holes in the $nf$ shell.
HCIs
have a number of important advantages: (i) the LV probe state is a ground state in many ions allowing for straightforward application of the scheme,
(ii) there is a wide variety of the ions to choose from, (iii) there is an extra enhancement factor with the degree of ionization.

The calculations for the monovalent ions are carried out using the linearized  coupled-cluster single-double
method (see \cite{SafJoh08} for a review). The calculation for the other ions are carried out using
 a method combining configuration interaction (CI)
with a modified linearized single-double coupled-cluster  approach~\cite{Koz04,SafKozJoh09}. The details of the calculations are
described in the supplemental material \cite{suppl}.
The results for selected HCIs are summarized in Table~\ref{ions}. We only list the HCIs where LV can be tested in the ground state since it simplifies the implementation scheme as it only requires a logic ion and RF pulses. The calculations are carried out for the ions already suggested for design of the atomic clocks and tests of $\alpha$ variation ~\cite{DzuDerFla12,SafDzuFla14,SafDzuFla14PRA1,SafDzuFla14PRA2,DzuSafSaf15}.
The table lists the reduced matrix elements $|\langle J ||T^{(2)}|| J \rangle|$ (in a.u.) and LLI-induced energy shift (in Hz)  between the
highest and lowest values of the magnetic quantum numbers $|m_J|$, for example $m_J=7/2$ and $m_J=1/2$  for $J=7/2$.  The Ca$^+$ and Yb$^+$ values are listed for reference. 
We list the number of the electrons $N$ for convenience. With the exception of the case with $N=58$, we only list the ions of the isoelectronic sequence with the lowest ionization charges which have at least one $nf$ electron in the ground state. More highly charged ions from the same isoelectronic sequence can be used  as well and are expected to have even larger sensitivities to LV. We demonstrate this point in the lower part of the table, where we list a number of ions with 58 electrons
and the same $4f^{12}$ ground state configurations but with increasing ionization charge.  Bi$^{25+}$, which can be produced with a small table-top electron-beam ion traps
already has factor of 4 larger matrix element in comparison with Yb$^+$. The enhancement with the ionization charge occurs for all other
isoelectronic sequences as well, so a very large number of HCIs is suitable for the LV tests using the experimental scheme describe above.
We also list Th$^{3+}$ since it can be directly laser cooled \cite{CamRadKuz11} and has $5f_{5/2}$ ground state. It can serve as excellent experiment test bed for later experiments with HCI.

In summary, we proposed an experimental scheme for drastic improvement of the LV tests in the electron sector. The scheme is applicable to any atomic spin system, including single and highly charged trapped ions and neutral atomic lattice clocks. It does not involve correlating operations between different spin probes, which simplify the experimental procedure to large extent.

This work was  supported in part by NSF grant PHY-1620687 (USA) and PHY-1507160
 (USA) and the UNSW group by the
Australian Research Council. R.S. and R.O. acknowledge support by the ICore-Israeli excellence center circle of light,
the Israeli Ministry of Science Technology and Space, the Minerva Stiftung
and the European Research Council (consolidator grant 616919-Ionology). M.S.S. thanks the School of Physics at UNSW, Sydney, Australia for hospitality
and acknowledges support from the Gordon Godfrey Fellowship program, UNSW.
S.G.P. acknowledges support from Russian Foundation for Basic Research under Grant No. 17-02-00216.

\begin{center}
\Large{\textbf{Supplemental Material}}\\
\end{center}
\section{Calculating the matrix elements}
\label{method}

The wavefunctions of the univalent ions are found in the framework of the coupled-cluster single double
method (see, e.g.,~\cite{BluJohLiu89}).
To find many-electron wave functions for divalent and trivalent ions we use a method combining configuration interaction (CI)
with a modified linearized single-double coupled-cluster (LCCSD) approach~\cite{Koz04,SafKozJoh09}.

At the CI stage we explicitly account for the interaction between valence electrons. The CI many-electron wave function can
be represented by a linear combination of the Slater determinants $\Phi_i$~\cite{KotTup87}:
\begin{equation}
\Psi = \sum_{i} c_{i} \Phi_i\,.
\end{equation}

We include the Breit interaction on the same footing as the Coulomb interaction at the stage of constructing the basis set, and incorporate the Gaunt part of the Breit interaction into the CI.

Further, we include core-valence correlations in the second order of the many-body perturbation theory (MBPT) over residual Coulomb
interaction~\cite{DzuFlaKoz96b} or using the LCCSD approach~\cite{Koz04,SafKozJoh09}, where the dominant core-valence correlations are included in all orders.

In both methods the one- and two-body parts $H_1$ and $H_2$ of the Hamiltonian $H$ are modified to include the correlation
potentials $\Sigma_1$ and $\Sigma_2$, correspondingly, that account for one- and two-body parts of the core-valence
correlations:
\begin{equation}  \label{H1eff}
H_k \rightarrow H_k+\Sigma_k,
\end{equation}
where $k=1,2$.

Then, the energies and wave functions of low-lying states are determined by
diagonalizing the effective Hamiltonian:
\begin{equation}
H^{\text{eff}}=H_{1} + H_2.
\label{ham}
\end{equation}

Such an approach allows us to improve the accuracy by an order of magnitude in comparison with a conventional CI method.
In the approach combining CI and LCCSD we include the dominant core-core and core-valence correlation corrections to the
effective Hamiltonian to all orders of the perturbation theory (we refer to it as the CI+all-order method)
that allows us further improve accuracy in comparison with the CI+MBPT method.
The detailed description of the CI+all-order method and all formulas are
given in~\cite{SafKozJoh09}. This method was successfully applied to calculation of the energy levels of Cd-like
and Sn-like ions~\cite{SafDzuFla14PRA2} and In-like ions~\cite{SafDzuFla14PRA1}.

The Ag-like ions are the univalent ions.
The Cd-like ions are divalent systems with two valence
electrons above the $[1s^2 2s^2 2p^6 3s^2 3p^6 3d^{10} 4s^2 4p^6 4d^{10}]$
core. The Sn-like ions, considered in this work, may be treated either
as divalent systems with the $[1s^2 2s^2 2p^6 3s^2 3p^6 3d^{10} 4s^2 4p^6 4d^{10} 5s^2]$ core or
systems with four valence electrons, when the $5s$ electrons are in the valence field.
Accounting for the fact that there are no low-lying states whose main configurations have the unpaired $5s$ electron, we are
considering these ions as the divalent ones in our calculations. In-like ions have the
core  $[1s^2 2s^2 2p^6 3s^2 3p^6 3d^{10} 4s^2 4p^6 4d^{10}]$ and three valence electrons above it. We consider them as the trivalent systems.

The neutral Cf and Es belong to the actinides group.
The Bi-like ions Cf$^{15+}$ and Es$^{16+}$ have the core $[1s^2, ..., 5d^{10}, 6s^2]$ and
three valence electrons above it. In this regard they resemble In-like ions but their valence
shells have the principle quantum number which is greater by 1 in comparison to the In-like ions.

Atomic states of Tm and Yb with open 4f shell are treated with the CIPT method~\cite{CIPT}. The correlations between 15 (for Tm) and 16 (for Yb) electrons on the outermost open shells are included.
The configuration interaction (CI) matrix is constructed for low-lying configurations while perturbation
theory is used to include higher configurations (see \cite{CIPT} for details).

Matrix elements of the LLI violating operator $\hat T$ are calculated with the use of the random-phase
approximation (RPA).  Core polarization is calculated self-consistently by solving the RPA equations for all
states in the core
\begin{equation}
(\hat H^{\rm HF} - \epsilon_a) \delta \psi_a = - (\hat T + \delta V^T)\psi_a.
\label{eq:RPA}
\end{equation}
Here $\hat H^{\rm HF}$ is the Hartree-Fock operator, index $a$ numerates states in the core, $\hat T$ is the
operator of the LLI violating external field, $ \delta V^T$ is the correction to the core  potential  caused by
external field. Matrix elements between valence states are calculated with the use of the modified operator
$\hat T + \delta V^T$.

One can also use the transitions between hyperfine structure sublevels of a term
to search for LLI violation. It increases the number of ions suitable for this purpose. In particular,
the ions with a nonzero nuclear spin $I$ whose ground states have $J \leq 1$ can also be used.
The recalcualtion $\langle J'||T^{(2)}||J \rangle$ to the hyperfine-coupled matrix element is given below.

In this case instead of the electron total angular momentum $J$ and its projections $m_J$ we need to use the total
angular momentum  $\mathbf{F}=\mathbf{J}+\mathbf{I}$
and its projections $M_F$. Using the Wigner-Eckart theorem and assuming that the tensor operator $T_0^{(2)}$ acts only to
the electronic part of the total wave function $|JIFM_F\rangle$, we have
\begin{eqnarray*}
\langle J'IF'M'_F|T_q^{(2)}|JIFM_F\rangle &=& (-1)^{F'-M'_F}
\left(
\begin{array}{ccc}
 F'   & 2 & F \\
-M'_F & q & M_F
\end{array}%
\right) \\
&\times&  \langle J'IF'||T^{(2)}||JIF \rangle ,
\end{eqnarray*}%
where%
\begin{eqnarray}
\langle J'IF'||T^{(2)}||JIF \rangle &=&(-1)^{F+J'+I}
\sqrt{(2F'+1)(2F+1)} \nonumber \\
&\times& \left\{
\begin{array}{ccc}
 J & I & F \\
F' & 2 & J'
\end{array}%
\right\} \langle J'||T^{(2)}||J \rangle .
\label{T2hfs}
\end{eqnarray}%
Thus,~\eref{T2hfs} gives the connection between $\langle J'IF'||T^{(2)}||JIF \rangle$ and
$\langle J'||T^{(2)}||J \rangle$ MEs.


\section{Expected energy resolution of the measurement scheme }

We now would like to estimate the precision $\Delta \kappa$ with
which we can hope to measure the energy shift $\kappa$ potentially containing the experimental signature of Lorentz violating effects.
Assuming that systematic drifts are absent, our measurement results can be described as the random variable $Y=\frac{1}{n}\sum_{i=1}^{n}x_{i}$, where $x_{i}$ are $n$ identical independent binomial distributed variables representing the state of the ions after  trial $i$.
$\Delta \kappa$, the change in $\kappa$ that can be observed between measurements is approximately given by
\begin{equation}
   \Delta \kappa= \left(\frac{d E[Y]}{d \kappa}\right)^{-1} \sqrt{V\left[E\left[Y\right]\right]},
\end{equation}
 where $E\left[Y\right]$ and $V\left[Y\right]$ denote the expectation value and variance of $Y$, respectively. The use of $\sqrt{V\left[E\left[Y\right]\right]}$ as a measure for the projection noise uncertainty assumes that $V\left[\sum_{i=1}^{n}x_{i}\right]\gg1$. Different uncertainty interval estimation should be used otherwise. In the proposed experiment, the probability of each spin to be detected in its initially prepared state is given by the $\kappa T$-dependent function $F\left(\kappa T\right)$, where $T$ is the Ramsey interrogation time. $x_{i}$ is a number between $0$ to $N$ corresponding to the number of spins detected in their initially prepared state, and therefore we use the expression $\sqrt{V\left[E\left[Y\right]\right]}=\sqrt{F\left(\kappa T\right)(1-F\left(\kappa T\right))N/n}$ and $E[Y]=NF\left(\kappa T\right)$ describing the expectation value as a function of $\kappa T$. We find
\begin{equation}
\Delta\kappa=\frac{\sqrt{F\left(\kappa T\right)\left(1-F\left(\kappa T\right)\right)}}{\sqrt{Nn}\frac{d}{d\kappa} F\left(\kappa T\right)}.
\end{equation}
choosing a specific measurement point $\kappa T = \chi$, the precision $\Delta \kappa$ is then given by
\begin{equation}
\Delta\kappa=\frac{\sqrt{F\left(\chi\right)\left(1-F\left(\chi\right)\right)}}{\sqrt{Nn}T\frac{d}{d\chi} F\left(\chi\right)}.
\end{equation}
Assuming that the total measurement time is dominated by the Ramsey interrogation time $T$, we can introduce the total measurement time $\tau=nT$ and obtain the expression for $\Delta \kappa$
\begin{equation}
\Delta\kappa=\frac{\sqrt{F\left(\chi\right)\left(1-F\left(\chi\right)\right)}}{\sqrt{N\tau T}\frac{d}{d\chi} F\left(\chi\right)}.
\end{equation}
For the proposed experiment, assuming no experimental phase drifts and considering $J=7/2$ we use $F\left(\chi\right)=P_{\frac{7}{2},m}\left(\chi,\pi\right)$. The calculated value of $\Delta \kappa$ for $\chi=\chi_{\mathrm{m}}$ corresponding to the maximal $\left|\frac{d}{d\chi} F\left(\chi\right)\right|$ is given by
\begin{eqnarray}
m=\frac{1}{2} \rightarrow \chi_{\mathrm{m}}\approx0.15\: \mathrm{rad},\: \Delta \kappa = 0.10\:\frac{\mathrm{rad}}{\sqrt{N\tau T}}\:,
\\
m=\frac{3}{2} \rightarrow \chi_{\mathrm{m}}\approx0.17\: \mathrm{rad},\: \Delta \kappa = {0.11}\:\frac{\mathrm{rad}}{\sqrt{N\tau T}}\:,
\\
m=\frac{5}{2} \rightarrow \chi_{\mathrm{m}}\approx0.20\: \mathrm{rad},\: \Delta \kappa = 0.17\:\frac{\mathrm{rad}}{\sqrt{N\tau T}}\:,
\\
m=\frac{7}{2} \rightarrow \chi_{\mathrm{m}}\approx0.22\: \mathrm{rad},\: \Delta \kappa = 0.28\:\frac{\mathrm{rad}}{\sqrt{N\tau T}}\:,\\
\end{eqnarray}
Here the working point $\chi'$ was chosen to decrease sensitivity to additional experimental noise that might reduce the contrast of $F\left(\chi\right)$, while maintaining reasonable uncertainty due to projection noise.

In order to compare these sensitivity factors to the equivalent sensitivity from Ref.~\cite{DzuFlaSaf16}, we use $F\left(\chi\right)=\frac{1}{2}\left(1+\sin\left[N\left(\left(\frac{7}{2}\right)^{2}-\left(\frac{1}{2}\right)^{2}\right)\chi\right]\right)$, where the $\frac{7}{2},\frac{1}{2}$ factors correspond to the choice of $m$ levels and $N$ is the (even) number of ions. This is a Ramsey spectroscopy signal obtained from a superposition in the form:
\begin{equation}
\frac{1}{\sqrt{2}}\left(\left|\frac{7}{2},\frac{7}{2}\right\rangle\left|\frac{7}{2},\text{-}\frac{7}{2}\right>^{\otimes\frac{N}{2}}+\left|\frac{7}{2},\frac{1}{2}\right\rangle\left|\frac{7}{2},\text{-}\frac{1}{2}\right>^{\otimes\frac{N}{2}}\right).
\end{equation}
For this experiment, we use $\sqrt{V\left[E\left[Y\right]\right]}=\sqrt{F\left(\chi\right)(1-F\left(\chi\right))/n}$ and $E[Y]=F\left(\chi\right)$ accounting for only two possible states for the value of each $x_{i}$. Assuming perfect Ramsey contrast, the precision becomes
\begin{equation}
\Delta \kappa = {0.083}\:\frac{\mathrm{rad}}{N\sqrt{\tau T}}\:.
\end{equation}

\section{Details of the experimental proof-of-concept}
\begin{figure}[h]
\begin{tabular}{cc}
\subf{\includegraphics[width=0.25\textwidth]{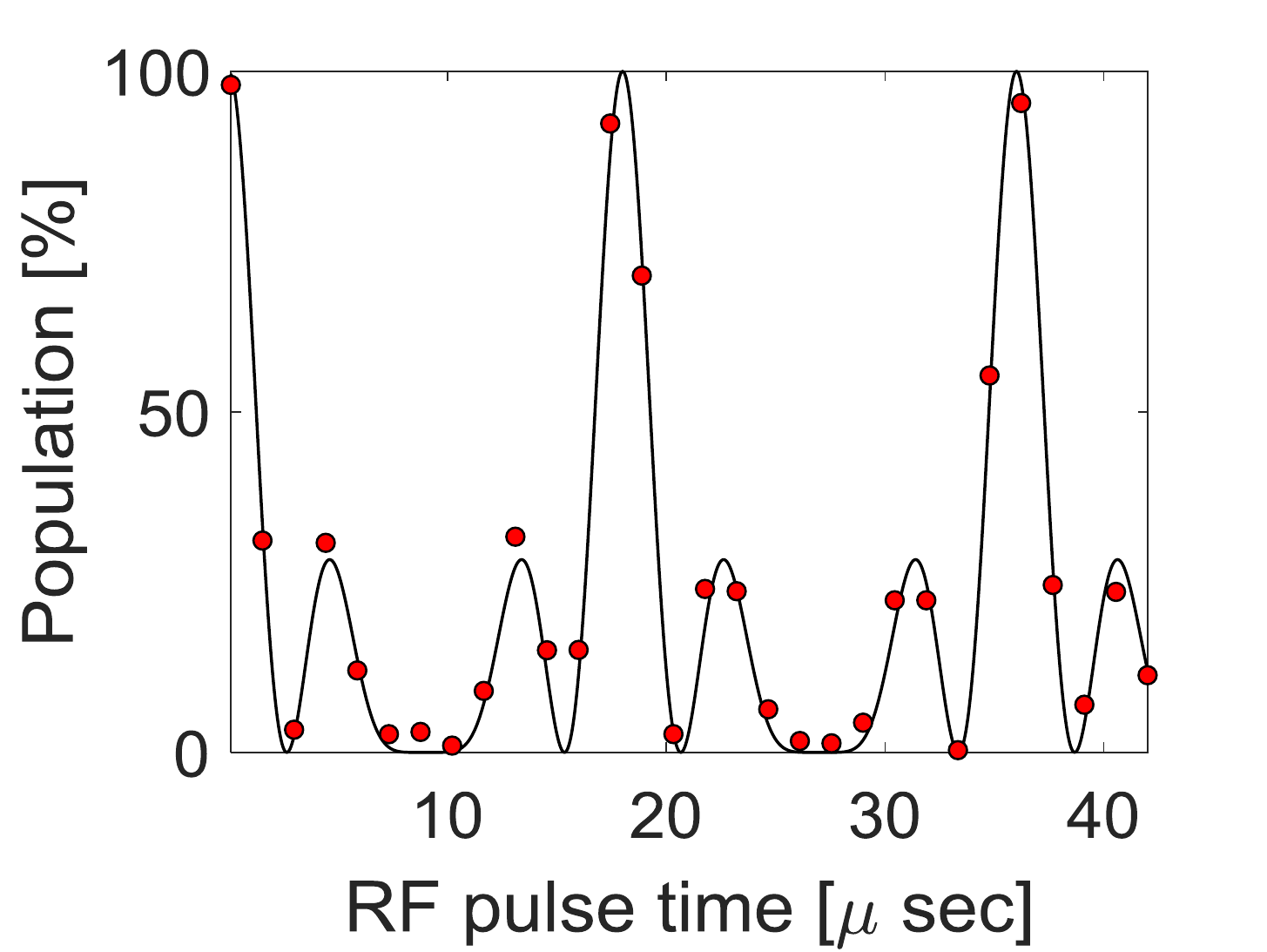}}{}
&
\subf{\includegraphics[width=0.25\textwidth]{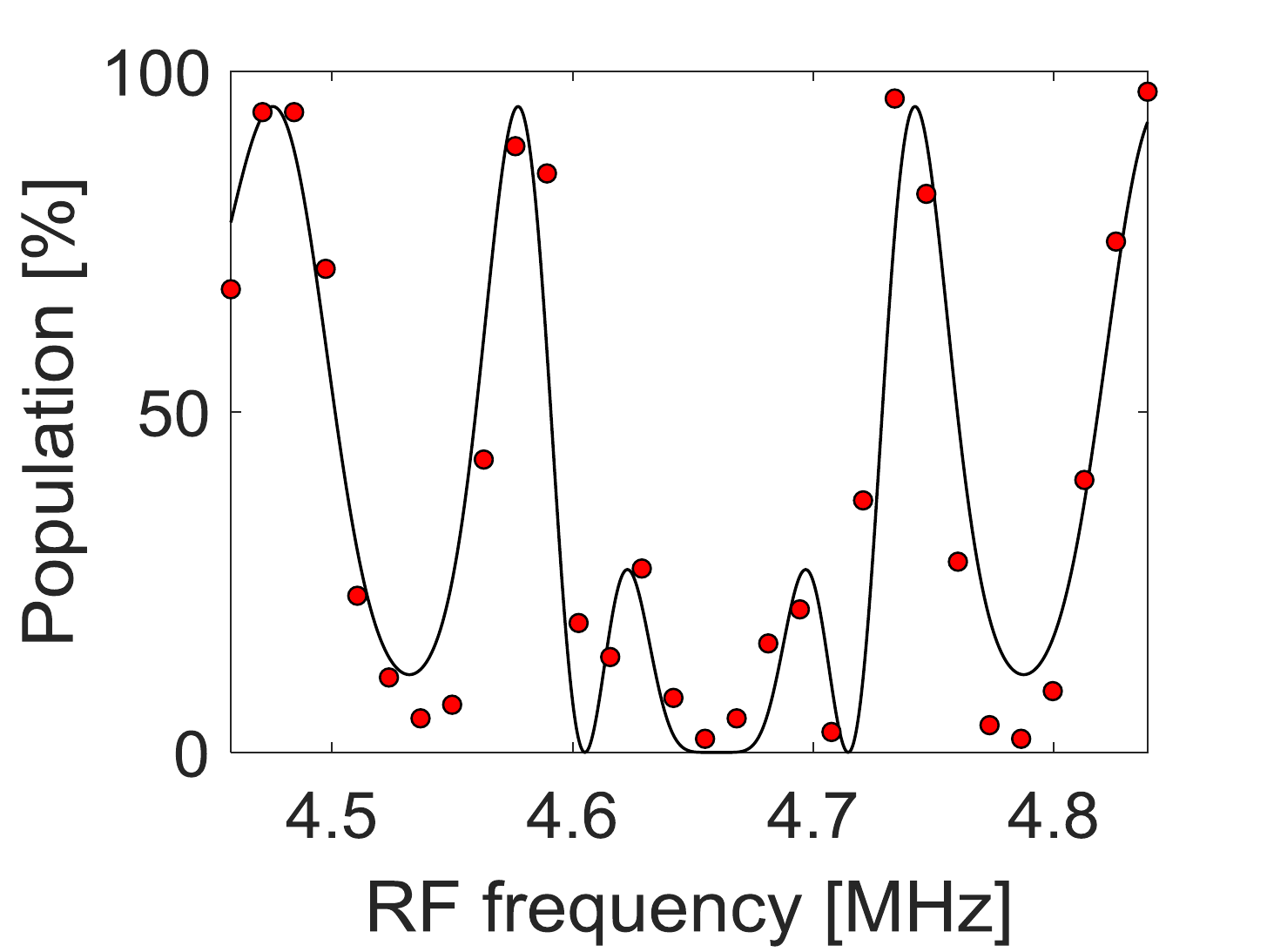}} {}
\end{tabular}
\caption{RF parameters calibration experiments results. Full red circles are experimental data and black solid curves are theoretical fit. \textbf{Left}: Rabi $\pi$ pulse time calibration. As in standard Rabi oscillations, in this calibration an RF on-resonance pulse time is scanned, and the population in the initial state is recorded. From the theoretical fit we extract the $\pi$ pulse time $t_{\pi}=\frac{\pi}{\Omega}$, where $\Omega$ is the RF Rabi frequency. Here we measured $t_{\pi}=8.99\, \mu \mathrm{sec}$, corresponding to $\Omega/2\pi\approx55\,\mathrm{kHz}$. \textbf{Right}: RF Resonance frequency calibration. Similar to Rabi spectroscopy experiment, the frequency of an RF pulse with duration time of $t_{\pi}$ is scanned and population in the desired state is measured. From the theoretical fit we extract the resonance frequency, $4.65 \, \mathrm{MHz}$.}
\label{RF_calibration}
\end{figure}

The experiment was done on a $^{88}\mathrm{Sr}^{+}$ two-ion chain trapped in a linear Paul trap \cite{wineland1998experimental}. The ions were initialized in the state $\left|4D_{\frac{5}{2}},m=-\frac{3}{2}\right>$ using an ultra-narrow linewidth laser at $674$ $\mathrm{nm}$ driving the quadrupole transition $\left|5S_{\frac{1}{2}},m_{\frac{1}{2}}=-\frac{1}{2}\right>\leftrightarrow\left|4D_{\frac{5}{2}},m=-\frac{3}{2}\right>$. Then, an oscillating current through a grounded electrode in the vicinity of the ions was used to implement the $J_{x},J_{y}$ operators for the DD sequence. Finally, the population in the initial state $\left|4D_{\frac{5}{2}},m=-\frac{3}{2}\right>$ was measured by mapping it to the $\left|5S_{\frac{1}{2}},m_{\frac{1}{2}}=-\frac{1}{2}\right>$ state and using state-selective fluorescence \cite{wineland1998experimental} induced by a $422$ $\mathrm{nm}$ laser driving the dipole transition $5S_{\frac{1}{2}}\leftrightarrow5P_{\frac{1}{2}}$. This fluorescence was detected by a fast EMCCD camera. More about the experimental apparatus can be found in Ref \cite{akerman2015universal}.  The RF resonance and Rabi frequency were measured with the same restrictions we impose in our DD scheme - only one state is allowed to be initialized and detected, with the right theory for the specific Rabi spectroscopy and Rabi oscillations. The calibration experiments results are shown in Fig. \ref{RF_calibration}.
During the measurement, active DC magnetic field compensation was operated, reducing the effect of slow magnetic field drifts. However, the magnetic noise induced by the AC line at frequency of 50 $\mathrm{Hz}$ and its harmonies was not compensated, and its amplitude was hundreds of $\mathrm{Hz}$ shift in the ion's RF resonance frequency. However, due to the DD scheme, as seen in Fig. 2 in the main text, the results follow the theoretical calculation with no magnetic noise. In addition, a loss of contrast can be seen from the comparison between theory and experimental data. This is mainly due to spontaneous decay from the $4D_{\frac{5}{2}}$ level to the ground state. The $4D_{\frac{5}{2}}$ lifetime is roughly $390\,\mathrm{msec}$, and therefore at $33\,\mathrm{msec}$ Ramsey time we would expect roughly $8.5\,\%$ loss of contrast.

\bibliography{ref1}

\end{document}